\pdfoutput=1

\documentclass[11pt]{article}

\usepackage[preprint]{acl}

\usepackage{times}
\usepackage{latexsym}

\usepackage[T1]{fontenc}

\usepackage[utf8]{inputenc}

\usepackage{microtype}

\usepackage{inconsolata}

\usepackage{graphicx}
\usepackage{amsmath}
\usepackage{booktabs}
\usepackage{enumitem}
\usepackage{multirow}
\usepackage{makecell}
\usepackage[normalem]{ulem}
\usepackage{acro}
\usepackage{paralist}
\usepackage{helvet}  
\usepackage{courier}  
\usepackage{natbib}  
\usepackage{caption} 
\usepackage[table]{xcolor}
\usepackage{algorithm}
\usepackage{algorithmic}
\usepackage{amssymb}
\usepackage{arydshln}

\DeclareAcronym{LLMs}{
  short = LLMs,
  long  = Large Language Models
}
\DeclareAcronym{RAG}{
  short = RAG,
  long  = Retrieval-Augmented Generation
}
\DeclareAcronym{GR}{
  short = GR,
  long  = Generative Retrieval
}
\DeclareAcronym{RL}{
  short = RL,
  long  = reinforcement learning
}
\DeclareAcronym{docids}{
  short = docids,
  long  = document identifiers
}

\title{Integrating Chain-of-Thought into Generative Retrieval: A Preliminary Study}

\author{
 \textbf{Wenhao Zhang\textsuperscript{1}},
 \textbf{Ruihao Yu\textsuperscript{1}},
 \textbf{Yi Bai\textsuperscript{1}},
 \textbf{Zhumin Chen\textsuperscript{1},
 \textbf{Pengjie Ren\textsuperscript{1}}\thanks{Corresponding author.}}
\\
 \textsuperscript{1}Shandong University, Qingdao, China
\\
 \texttt{\{zhangwenhao, 202322130199, 202235147\}@mail.sdu.edu.cn,} \\
 \texttt{\{chenzhumin,renpengjie\}@sdu.edu.cn}
}

\begin{document}
\maketitle
\begin{abstract}
  While generative retrieval (GR) demonstrates competitive performance on standard retrieval benchmarks, existing approaches directly map queries to document identifiers (docids) without intermediate deliberation, limiting their effectiveness for complex queries that require multi-step reasoning.
As a preliminary study on integrating chain-of-thought (CoT) into generative retrieval, we introduce ThinkGR, a unified framework that interleaves CoT with docid generation, enabling iterative thinking and retrieval within a single generative process.
To bridge the gap between free-form thought generation and structured retrieval targets, we design (1) a hybrid decoding strategy that dynamically switches between unconstrained thought generation and constrained docid decoding, and (2) a two-phase training approach that first aligns thought-retrieval patterns through supervised fine-tuning, then optimizes thought quality via retrieval-grounded reinforcement learning.
Experiments on four multi-hop retrieval benchmarks demonstrate that ThinkGR achieves state-of-the-art performance with an average improvement of +6.86\%.
Our work opens new avenues for enhancing generative retrieval with explicit deliberation capabilities, with promising implications for retrieval tasks requiring complex reasoning.
\end{abstract}

\section{Introduction}

\begin{figure}[t!]
    \centering
    \includegraphics[width=\columnwidth]{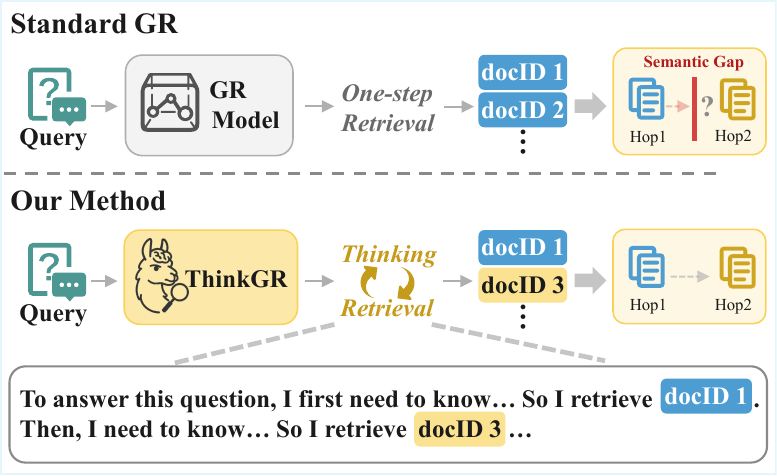}
    \caption{Comparison between standard generative retrieval and our thought-augmented approach.
    (a) Standard GR: Directly generates docids from the query without intermediate deliberation.
    (b) ThinkGR: Interleaves chain-of-thought with docid generation, enabling iterative thinking and retrieval within a single forward pass.
    }
    \label{fig:intro}
\end{figure}

\ac{GR} has emerged as a promising paradigm that reformulates document retrieval as a sequence generation task~\cite{deautoregressive, tay2022transformer, bevilacqua2022autoregressive}.
Unlike traditional dense retrieval methods that rely on embedding similarity matching, \ac{GR} directly generates \ac{docids} from queries by leveraging generative model architectures.
This approach enables end-to-end optimization and has demonstrated competitive performance on standard retrieval benchmarks~\cite{zhou2022ultron, zhang2024generative, NEURIPS2023_91228b94}.
Recent work by \citet{zhang2025does} reproduced representative DR and GR under matched backbones, and through controlled and transparent comparisons, showed that \ac{GR} surpasses standard dual-encoder dense retrievers on Natural Questions~\cite{kwiatkowski2019natural}.

Despite these advances, existing \ac{GR} methods share a fundamental limitation: they directly map queries to \ac{docids} without intermediate deliberation.
This single-step generation paradigm works well for straightforward queries where the target document is semantically close to the query surface form.
However, when queries require complex reasoning, such as in multi-hop retrieval scenarios where resolving the query necessitates traversing semantic connections across multiple documents~\cite{yang-etal-2018-hotpotqa, trivedi2022musique}, the lack of an explicit thinking process severely limits retrieval effectiveness.
For example, given the query ``\textit{What is the capital of the country where the director of Erta Ale was born?}'', answering it requires first retrieving documents about the director of Erta Ale, then retrieving documents about the corresponding country's capital---a scenario where standard single-hop retrieval models struggle to retrieve the final target document, as it lacks direct semantic similarity to the original query.

Inspired by recent advances in chain-of-thought (CoT) prompting~\cite{wei2022chain} and deliberative reasoning models~\cite{jaech2024openai, guo2025deepseek}, we investigate enabling \ac{GR} to interleave reasoning and retrieval within a unified generative process.
This iterative paradigm can bridge the semantic gap between complex queries and target documents through intermediate reasoning steps (Figure~\ref{fig:intro}).
While recent concurrent work has begun to explore reasoning in generative retrieval, existing approaches either limit reasoning to a single pre-retrieval step~\cite{dong2026multi} or achieve iterative reasoning through separate sequential inference calls without end-to-end training~\cite{zhang2025retrieval}.
However, effectively merging the thought process into retrieval models within a single end-to-end generative pass remains challenging due to the inherent differences between open-ended reasoning and discrete retrieval actions.

To address this challenge, we propose ThinkGR, a unified framework that integrates CoT into \ac{GR}, enabling the model to iteratively think and retrieve within a single generative process.
To unify the decoding space, we design a hybrid decoding strategy that dynamically switches between unconstrained generation for thought tokens and constrained decoding for \ac{docids}, using semantic triples as document representations that naturally bridge natural language and structured retrieval targets.
To unify the learning process, we employ a two-phase training strategy: first aligning thought-retrieval patterns through supervised fine-tuning, then optimizing thought quality using retrieval accuracy as a grounded reward signal.

We conduct extensive experiments on four multi-hop retrieval benchmarks: HotpotQA, 2WikiMultihopQA, MuSiQue, and MoreHopQA.
Results demonstrate that ThinkGR achieves state-of-the-art performance, significantly outperforming existing methods with an average improvement of +6.86\%.

Our contributions can be summarized as follows:
\begin{inparaenum}[(1)]
\item We present ThinkGR, a unified framework that integrates chain-of-thought into generative retrieval by jointly generating iterative thought and retrieval sequences within a single end-to-end trainable autoregressive pass, demonstrating the feasibility and effectiveness of this paradigm for complex queries.
\item We identify key challenges in enabling iterative thinking and retrieval, and propose solutions including a hybrid decoding strategy and a two-phase training approach.
\item ThinkGR achieves state-of-the-art results on four benchmarks, demonstrating its effectiveness in multi-hop retrieval scenarios that require complex thinking.
\end{inparaenum}

\section{Related Work}

\subsection{Thought-Augmented Retrieval}

Recent advances in chain-of-thought prompting~\cite{wei2022chain} and deliberative reasoning~\cite{jaech2024openai, guo2025deepseek} have inspired efforts to incorporate thought processes into retrieval systems.

\textbf{LLM-Driven Multi-Step Retrieval.}
One line of work leverages LLMs' intrinsic reasoning capabilities to decompose complex queries and guide iterative retrieval.
ReAct~\cite{yao2023react} alternates between reasoning steps and retrieval actions;
Self-ask~\cite{press-etal-2023-measuring} decomposes questions into sub-questions with follow-up Q\&A;
IRCoT~\cite{trivedi-etal-2023-interleaving} interleaves retrieval with stepwise reasoning to refine subsequent searches.
Recent methods employ reinforcement learning: Auto-RAG~\cite{yu2024autorag} trains LLMs to decide retrieval timing; DeepRAG~\cite{guan2025deeprag} models the process as a Markov Decision Process; R3-RAG~\cite{li2025r3} uses outcome and process rewards for optimization.
However, these approaches require separate LLM and retriever modules with iterative handoffs, preventing end-to-end optimization.

\textbf{Reasoning-Augmented Dense Retrieval.}
Another line integrates thought processes into dense retrieval models.
Early methods incorporate thought implicitly without explicit tokens: MDR~\cite{xiong2020answering} iteratively encodes queries with retrieved context to guide subsequent retrieval; GRITHopper~\cite{erker2025grithopper} unifies language modeling and contrastive retrieval in a single LLM; O1 Embedder~\cite{yan2025o1} leverages LLM-generated ``thoughts'' as intermediate representations before aggregating into embeddings.
Recent work explores explicit thought tokens before embedding generation: GEM~\cite{zhang2025gem} inserts bottleneck tokens with specialized attention masks to compress semantic information; LREM~\cite{tang2025large} establishes a ``think-then-embed'' paradigm where models generate keyword-form CoT before final embeddings.
While these methods demonstrate the value of explicit thought for retrieval, they operate within dense retrieval and typically perform thought generation only once before embedding.
In contrast, our work integrates thought into generative retrieval, enabling \textit{iterative} thought-retrieval interleaving within a single end-to-end decoding pass, particularly beneficial for multi-hop scenarios requiring progressive reasoning.

\subsection{Generative Retrieval}

Generative retrieval formulates document retrieval as a sequence generation task, directly producing \ac{docids} through autoregressive decoding~\cite{deautoregressive, bevilacqua2022autoregressive, NEURIPS2023_91228b94}.
Previous research has explored various types of \ac{docids}, which can be categorized into lexical ids and numeric ids.
Lexical ids include title~\citep{deautoregressive}, substring~\citep{bevilacqua2022autoregressive}, URL~\citep{zhou2022ultron}, term-sets~\citep{zhang2024generative}, etc., while numeric ids are typically obtained through clustering of document representations~\citep{tay2022transformer, zhou2022ultron, NEURIPS2023_91228b94}.
This paradigm enables end-to-end optimization and has shown competitive performance on standard benchmarks~\cite{zhang2025does, zhang2025excluir}.
While \ac{GR} has achieved promising results, existing approaches lack intermediate deliberation mechanisms, limiting their effectiveness on complex queries.
Although \citet{lee-etal-2022-generative} applied \ac{GR} to multi-hop settings using document fragments, their approach lacks explicit thought mechanisms and faces scalability challenges for large corpora.
\citet{li2024corpuslm} propose CorpusLM, a unified language model that integrates \ac{GR} with retrieval-augmented generation through a continuous decoding process, where the model sequentially generates \ac{docids} followed by document references and final answers.
While CorpusLM demonstrates the potential of unifying retrieval and generation, free-text generation occurs after retrieval to produce answers rather than interleaving explicit reasoning during the retrieval process itself.

Recent work has begun to explore incorporating reasoning directly into the generative retrieval pipeline.
\citet{dong2026multi} propose ReasonGR, which augments an encoder-decoder GR model with chain-of-thought-style prompting for complex numerical reasoning over financial documents, improving single-step docid generation without iterative thought-retrieval interleaving.
\citet{liu2025onerec} extend this idea to generative recommendation with OneRec-Think, generating interpretable reasoning paths before producing item identifiers in a single pass.
Most closely related to our work, \citet{zhang2025retrieval} propose Retrieval-in-the-Chain (R4R), which converts free-form CoT into compact structured signals and performs iterative retrieve-refine cycles within a GR framework.
However, R4R achieves this iterative process through separate sequential inference steps that include think, retrieve, verify, and reflect, requiring multiple LLM calls per query.
It does not employ end-to-end training to jointly optimize reasoning and retrieval, relying instead on prompt-based reasoning at inference time.
Furthermore, R4R is primarily validated on single-hop retrieval benchmarks.
In contrast, ThinkGR generates iterative thought-retrieval sequences within a single autoregressive decoding pass, with end-to-end retrieval-grounded training that jointly optimizes thought quality and retrieval accuracy, and is specifically designed and evaluated for complex multi-hop scenarios.

\begin{figure*}[ht]
    \centering
    \includegraphics[width=2\columnwidth]{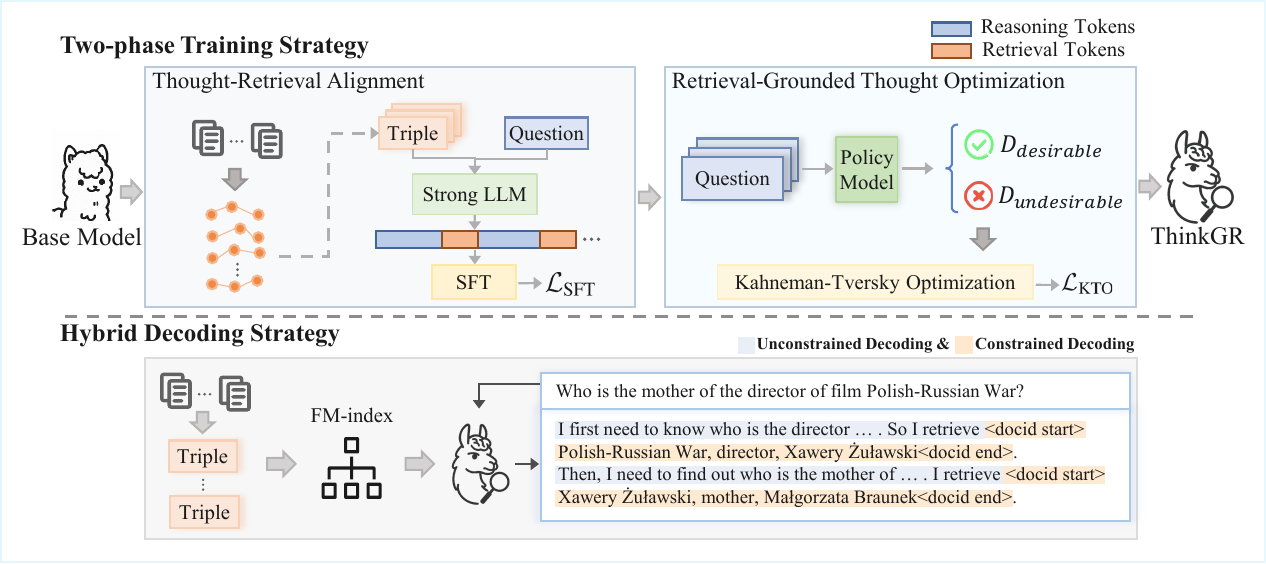}
    \caption{Overview of the ThinkGR framework.
    Top: Two-phase training strategy (Thought-Retrieval Alignment followed by Retrieval-Grounded Thought Optimization).
    Bottom: Hybrid Decoding Strategy combining constrained and unconstrained decoding.}
    \label{fig:overview}
\end{figure*}



\section{Method}

To realize the idea of integrating chain-of-thought into generative retrieval, we develop ThinkGR as a preliminary instantiation.
ThinkGR formulates retrieval as a unified sequence generation problem, generating an interleaved sequence of thought tokens and \ac{docids} through a single forward pass, where each thought segment contextualizes and guides the subsequent retrieval decision.
We implement ThinkGR by training Llama-3.1-8B-Instruct with a two-phase training strategy.
Figure~\ref{fig:overview} shows an overview of our method.

\subsection{Semantic Triple Representation}

A critical design choice in generative retrieval is how to represent documents as generation targets.
Traditional docids fail to capture the relational semantics essential for complex queries.
We observe that multi-hop queries inherently require traversing semantic relationships between entities, a structure naturally expressed as knowledge triples.

We thus represent documents as structured knowledge triples (head entity, relation, tail entity), serving as \ac{docids}.
This design serves two purposes: (1) triples encode fine-grained semantic relationships, enabling the model to perform semantic traversal through explicit relation following, and (2) natural-language triples allow the model to leverage the pre-trained LLM's semantic understanding to generalize to unseen entities and relations.
We design a prompt to instruct LLM to generate these triples.
To construct SFT data, we further prompt the LLM to generate thought-retrieval chains based on the correct document triples and questions.
Detailed prompts are provided in Appendix~\ref{appendix:prompt}.
To ensure high-quality training data and minimize cascading errors, we implemented a rigorous filtering process that removes samples with formatting errors, incorrect triples, and factual inaccuracies.
This process resulted in a high-quality curated set of 228K SFT data from HotpotQA, 2WikiMultihopQA, and Musique datasets.

\subsection{Thought-Retrieval Alignment}

The first training phase establishes the model's ability to generate interleaved thought-retrieval sequences.
By focusing on this structural alignment, we enable the model to utilize its pre-trained knowledge for robust generalization to unseen entities and relations, even if they were not explicitly seen during fine-tuning.
In this process, the generated thought steps naturally guide the retrieval operations, forming a coherent and effective workflow.
Formally, given input query $x$, the model learns to generate an interleaved sequence $y = (r_1, d_1, r_2, d_2, \ldots)$ where $r_i$ denotes thought segments and $d_i$ denotes \ac{docids}.
The training objective minimizes the negative log-likelihood:
\begin{equation}
\mathcal{L}_{\text{SFT}} = -\sum_{t} \log P(y_t | y_{<t}, x),
\end{equation}
where $y_t$ denotes each token in the output sequence.
After this phase, the model acquires the basic thought-retrieval generation capability, preparing it for further optimization.

\subsection{Retrieval-Grounded Thought Optimization}

While the alignment phase teaches the model the structural pattern of thought-retrieval interleaving, it is limited by the quality of demonstration data and cannot discover better thought strategies beyond imitation.
To address this, we introduce a retrieval-grounded reinforcement learning phase that uses retrieval accuracy as a proxy reward for thought quality.
The key insight is that in our unified framework, the quality of thought directly manifests in retrieval outcomes.
This creates a natural feedback loop: we can use retrieval accuracy as an automatic and grounded reward signal to optimize the thought process, without requiring expensive human annotations of thought quality.

We implement this using Kahneman-Tversky Optimization (KTO)~\cite{ethayarajh2024kto}.
Unlike PPO~\cite{schulman2017proximal} or DPO~\cite{rafailov2023direct} which require preference pairs, KTO operates effectively with binary feedback, allowing us to directly leverage retrieval correctness as supervision.
Specifically, if the generated \ac{docids} match the ground-truth exactly, the response is labeled as \textit{desirable}; if the accuracy of generated \ac{docids} is less than $\tau$, it is labeled as \textit{undesirable}.
This creates a grounded optimization objective that directly ties reasoning quality to task performance.
Formally, we partition model-generated responses based on retrieval accuracy:
\begin{align}
D_{\text{desirable}} &= \{(x, y) | \text{Acc}_\text{r}(y) = 1\}, \\
D_{\text{undesirable}} &= \{(x, y) | \text{Acc}_\text{r}(y) < \tau\}, \\
\text{Acc}_\text{r}(y) &= \frac{|\text{docids}(y) \cap \text{docids}_\text{gr}|}{|\text{docids}_\text{gr}|},
\end{align}
where $\text{docids}(y)$ denotes the \ac{docids} generated by the model, and $\text{docids}_\text{gr}$ denotes the ground-truth \ac{docids}.
After filtering, we obtained 60K desirable responses and 27K undesirable responses.
KTO can implicitly handle the data imbalance through adaptive weighting, so we did not perform further balanced sampling.
We optimize the model weights based on the following loss function:
\begin{equation}
\begin{split}
\mathcal{L}_{\text{KTO}} = &\mathbb{E}_{(x, y) \sim D_{\text{desirable}}} [\lambda_d - v(x, y)] \\
+ &\mathbb{E}_{(x, y) \sim D_{\text{undesirable}}} [\lambda_u - v(x, y)],
\end{split}
\end{equation}
where
{\small
\begin{align*}
    r_\theta(x, y) &= \log{\frac{\pi_\theta(y|x)}{\pi_\text{ref}(y|x)}} \\
    z_0 &= KL(\pi_\theta(y'|x) || \pi_\text{ref}(y'|x)) \\
    v(x, y) &= 
    \begin{cases} 
    \lambda_d(\beta(r_\theta(x, y) - z_0)) & \text{if } y \in D_{\text{desirable}} \\
    \lambda_u(\beta(z_0 - r_\theta(x, y))) & \text{if } y \in D_{\text{undesirable}},
    \end{cases}
\end{align*}}
where $v(x,y)$ represents the prospect-theoretic utility function that converts the model's implicit reward into a human-perceived value relative to a reference point.
$\pi_\theta$ refers to the policy model being trained, where $\pi_\theta(y|x)$ denotes the probability of generating output sequence $y$ given input $x$ under parameters $\theta$.
$\pi_\text{ref}$ represents the reference model, which is the supervised fine-tuned model before reinforcement learning.
$r_\theta(x, y)$ is the implied reward, $z_0$ is the reference point dynamically estimated per mini-batch, $\lambda_d$ and $\lambda_u$ are automatically adjusted parameters for data imbalance, and $\beta$ is a risk aversion hyperparameter controlling optimization sensitivity.

\subsection{Hybrid Decoding Strategy}

The unified sequence formulation introduces a unique inference challenge: the model must generate tokens from two different spaces\textemdash an open vocabulary for thought and a constrained set of valid \ac{docids} for retrieval\textemdash within a single autoregressive process.
Naive unconstrained decoding risks generating invalid docids (hallucinations), while fully constrained decoding would prevent flexible thought generation.

We address this through a hybrid decoding strategy that dynamically switches between unconstrained and constrained generation modes.
For constrained docid generation, we employ FM-index~\cite{ferragina2000opportunistic} following \citet{bevilacqua2022autoregressive}, which returns valid next tokens in constant time.
We preprocess corpus triples into the format ``\textless docid\_start\textgreater head entity, relation, tail entity\textless docid\_end\textgreater'' and store them in the FM-index.
During inference, the decoding process operates as follows:
(1) ThinkGR begins with unconstrained decoding to generate thought tokens, enabling ``slow thinking'' where the model elaborates on the query context and determines what information is needed;
(2) When the model decides a retrieval is necessary, it generates ``\textless docid\_start\textgreater'', which triggers a switch to constrained decoding mode;
(3) In constrained mode, the FM-index restricts the output vocabulary to valid next tokens based on the current prefix, ensuring the generated docid exists in the corpus;
(4) Upon generating ``\textless docid\_end\textgreater'', the mode switches back to unconstrained for continued thought.

This hybrid strategy realizes the core idea of thought-augmented retrieval: the model autonomously decides when to retrieve and what to retrieve through explicit deliberation, all within a single autoregressive pass.
This design demonstrates the feasibility of unifying free-form thought with constrained retrieval generation, eliminating the latency overhead of iterative LLM-retriever invocations while maintaining retrieval validity.

\section{Experiments}

\begin{table*}[ht]
  \centering
  \begin{tabular}{lcccccc}
    \toprule
    \textbf{Method} & \textbf{Model Parameters} & \textbf{HotpotQA} & \textbf{2Wiki.} & \textbf{Musique} & \textbf{Morehopqa} & \textbf{Average} \\
    \hline
    \rowcolor{gray!25}
    \multicolumn{3}{l}{\textit{Standard Retriever}} & & & & \\
    BM25 & - & 46.39 & 49.80 & 31.57 & 43.74 & 42.88 \\
    Contriever & - &  50.35 & 51.25 & 34.02 & 45.04 & 45.17 \\
    BGE-large & 326M &  60.48 & 58.43 & 33.39 & 47.58 & 49.97 \\
    SEAL & 406M & 56.15 & 48.32 & 35.60 & 47.27 & 46.84 \\
    \hdashline
    \rowcolor{gray!25}
    \multicolumn{3}{l}{\textit{LLM-Driven Multi-Step Retrieval}} & & & & \\
    Selfask & 70B & 44.40 & 47.07 & 34.28 & 57.60 & 45.84 \\
    IRCoT & 70B & 55.79 & 65.12 & 49.96 & 66.82 & 59.42 \\
    ITER-RETGEN & 70B & 61.94 & 59.96 & 39.37 & 60.73 & 55.50 \\
    Auto-RAG & 7B & 54.52 & 66.05 & 40.63 & 59.48 & 55.17 \\
    R3-RAG & 8B & 58.56 & \underline{82.82} & 51.70 & 70.44 & 65.88 \\
    RT-RAG & 8B & 62.48 & 65.14 & 51.30 & 66.95 & 61.47 \\
    \hdashline
    \rowcolor{gray!25}
    \multicolumn{3}{l}{\textit{Reasoning-Augmented Dense Retrieval}} & & & & \\
    MDR & 110M & \underline{87.57} & 53.91 & 27.84 & 49.60 & 54.73 \\
    GritHopper & 7B & \textbf{91.03} & 59.97 & \underline{60.48} & \underline{74.82} & \underline{71.58} \\
    \hdashline
    \rowcolor{gray!25}
    \multicolumn{3}{l}{\textit{Our Method}} & & & & \\
    ThinkGR & 8B & 76.09 & \textbf{93.19} & \textbf{63.98} & \textbf{80.50} & \textbf{78.44} \\
    \textit{w/o} SFT & 8B & 35.76 & 57.70 & 35.98 & 55.14 & 46.15 \\
    \textit{w/o} RL & 8B & 67.66 & 92.03 & 53.40 & 73.84 & 71.73 \\
    \textit{w/o} Thought & 8B & 69.86 & 93.01 & 53.23 & 78.00 & 73.53 \\
    \bottomrule
  \end{tabular}
  \caption{Performance comparison of ThinkGR with baselines on four datasets.
  The results are reported in terms of recall.
  The best results for each dataset are highlighted in \textbf{bold}, and the second-best results are \underline{underlined}.
  The model parameters indicate the number of parameters in millions (M) or billions (B).}
  \label{tab:main_results}
\end{table*}

\subsection{Datasets and Metrics}
\label{sec:datasets}
Since our method is specifically designed for retrieval scenarios requiring complex reasoning, we evaluate its effectiveness on four commonly used multi-hop QA datasets: \textbf{HotpotQA}~\cite{yang-etal-2018-hotpotqa}, \textbf{2WikiMultiHopQA}~\cite{ho-etal-2020-constructing}, \textbf{MuSiQue}~\cite{trivedi2022musique}, and \textbf{MoreHopQA}~\cite{schnitzler2024morehopqa}.
Detailed descriptions and statistics are provided in Appendix~\ref{appendix:datasets}.

We use \textit{Recall} as our primary evaluation metric, defined as the ratio of correct documents retrieved to the total number of ground-truth documents.
This metric is a common choice in multi-hop retrieval benchmarks~\cite{trivedi-etal-2023-interleaving, erker2025grithopper}, allowing for direct and fair comparison with prior work.

\subsection{Baselines}

We compare our method with standard one-hop retrievers, including \textit{BM25}~\cite{robertson2009probabilistic}, \textit{Contriever}~\cite{izacard2021contriever}, \textit{BGE}~\cite{bge_embedding}, and generative retrieval method \textit{SEAL}~\cite{bevilacqua2022autoregressive}.
To comprehensively assess multi-hop retrieval capability, we primarily compare with two categories of methods specifically designed for this task: LLM-driven multi-step retrieval and reasoning-augmented dense retrieval.
LLM-driven multi-step retrieval methods include: \textit{Self-Ask}~\cite{press-etal-2023-measuring}, prompts the LLM to generate follow-up questions based on the current context; \textit{IRCoT}~\cite{trivedi-etal-2023-interleaving}, uses CoT generated by LLM to guide retrieval; \textit{ITER-RETGEN}~\cite{shao-etal-2023-enhancing}, iteratively performs reasoning and retrieval; \textit{Auto-RAG}~\cite{yu2024autorag}, employs reinforcement learning to enable autonomous decision-making on retrieval through multi-turn LLM-retriever interactions; \textit{R3-RAG}~\cite{li2025r3}, optimizes iterative retrieval via reinforcement learning with outcome and process rewards for document relevance verification; and \textit{RT-RAG}~\cite{shi2026reasoning}, constructs explicit hierarchical reasoning trees by decomposing multi-hop questions into structured sub-queries and employs a bottom-up traversal strategy with iterative query rewriting to collect evidence.
Reasoning-augmented dense retrieval methods include: \textit{MDR}~\cite{xiong2020answering} iteratively encodes the concatenated question and previously retrieved documents into a single vector for next-hop retrieval;
\textit{GRITHopper}~\cite{erker2025grithopper} integrates causal language modeling with contrastive dense retrieval through ReAct-style instruction tuning, leveraging post-retrieval language modeling to contextualize multi-hop reasoning.
Implementation details are provided in Appendix~\ref{appendix:implementation}.

\section{Experimental Results}

We evaluate ThinkGR on four multi-hop QA benchmarks to demonstrate the effectiveness of integrating chain-of-thought into generative retrieval.

\subsection{Main Results}
\label{sec:main_results}

The main results on the four datasets are presented in Table \ref{tab:main_results}.
We derive the following observations from the results:

(1) Our method demonstrates state-of-the-art performance, achieving an average performance gain of 6.86\% over the strongest baseline.
This establishes the effectiveness of interleaving iterative thinking with retrieval in a unified generative process.
Note that while ThinkGR does not achieve the best result on HotpotQA, this is primarily because HotpotQA suffers from the over-specification issue.
This allows models to find relevant documents through simple lexical matching, making it unfair to evaluate the thought-retrieval capability of models.
To illustrate this point, we conducted a statistical analysis of the n-gram overlap between questions and ground-truth documents across three datasets.
The results indicate that HotpotQA indeed exhibits a significantly high overlap rate.
Detailed statistical results are presented in Appendix~\ref{appendix:over_specification}.
Despite this, ThinkGR still significantly outperforms all LLM-Driven Multi-Step Retrieval methods on HotpotQA.

(2) ThinkGR exhibits superior generalization ability on out-of-domain evaluation.
Compared to GritHopper, ThinkGR achieves a 5.68\% improvement on the out-of-domain dataset Morehopqa.
This indicates that ThinkGR is more robust and can generalize better to unseen data and schema not encountered during training, which is crucial for real-world applications.

(3) ThinkGR shows better stability across different datasets.
It is evident that Reasoning-Augmented Dense Retrieval methods generally outperform LLM-Driven Multi-Step Retrieval methods, but they perform poorly on the 2WikiMultihopQA dataset.
ThinkGR achieves state-of-the-art results on both 2WikiMultihopQA (simpler queries) and Musique (more complex queries), demonstrating strong generalization across different difficulty levels.
We conducted a statistical analysis and found that GritHopper retrieves an average of 1.79 documents on 2WikiMultihopQA, while the ground-truth documents average 2.44.
This indicates that GritHopper retrieves insufficient documents on 2WikiMultihopQA, leading to its suboptimal performance.
In contrast, our method ThinkGR retrieves an average of 2.85 documents on 2WikiMultihopQA, further demonstrating the better stability of our method across different datasets.

\begin{figure}
    \centering
    \includegraphics[width=\columnwidth]{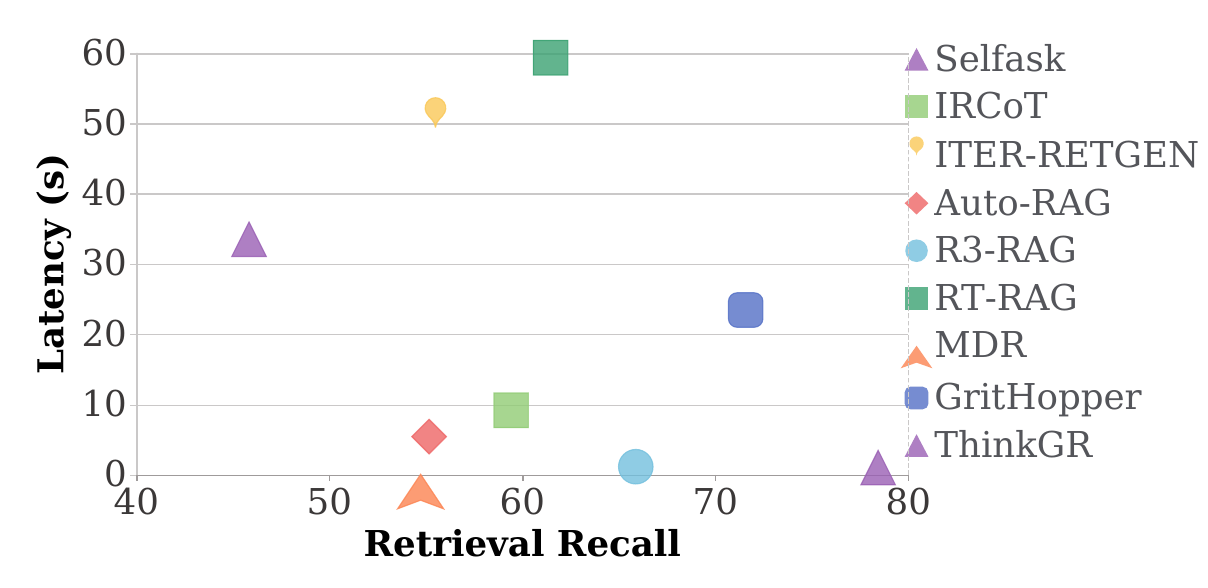}
    \caption{Comparison of effectiveness and efficiency.
    The x-axis denotes retrieval recall and the y-axis represents average latency per query.}
    \label{fig:speed}
\end{figure}

\subsection{Ablation Studies}
\label{sec:ablation}

\paragraph{Effectiveness of Two-Phase Training.}
\label{sec:ablation_train}
We investigate the effectiveness of the two-phase training strategy in ThinkGR.
The results of models trained only with thought-retrieval alignment or retrieval-grounded thought optimization are presented in Table~\ref{tab:main_results}.
From the results, we draw two observations.
First, the model trained without retrieval-grounded thought optimization exhibits a significant drop in retrieval performance.
This indicates that this optimization phase substantially enhances the model's thought capabilities, enabling it to surpass the performance ceiling achieved by SFT.
We illustrate this point further with a specific case study in Appendix~\ref{appendix:case_study}.
Second, relying solely on retrieval-grounded thought optimization is insufficient for the model to learn the thought-retrieval workflow.
This is evidenced by the overall inferior performance of models trained without prior thought-retrieval alignment.
This suggests that supervised training is necessary to adapt the model to the retrieval task.

\paragraph{Effectiveness of Hybrid Decoding.}
\label{sec:ablation_reason}
We next ablate the hybrid decoding strategy\textemdash the interleaved generation of natural language thought tokens and \ac{docids} within a single output sequence.
To evaluate the contribution of this iterative thinking mechanism, we train a variant that generates only the \ac{docids} sequence, omitting the thought tokens entirely.
Essentially, this transforms the explicit thought into implicit thought within the model parameters.
The results in Table~\ref{tab:main_results} show that this docids-only variant (w/o Thought) reduces retrieval accuracy, especially on the more challenging dataset Musique.
This variant essentially represents a standard constrained decoding strategy where the model directly generates retrieval targets without intermediate thought.
This indicates that the interleaved thought tokens are not merely interpretative, but also crucial for bridging semantic gaps across retrieval steps in complex queries.
Thus, this ablation study demonstrates that unifying iterative thinking and retrieval into a single generation is superior to methods that rely on implicit reasoning like GritHopper or our docids-only variant.
This unified hybrid decoding strategy is essential to ThinkGR, as it enables complex, iterative thought and retrieval in a single end-to-end pass.

\begin{table*}[ht]
  \centering
  \begin{tabular}{lccccc}
    \toprule
    \textbf{Model} & \textbf{HotpotQA} & \textbf{2WikiMultihopQA} & \textbf{Musique} & \textbf{Morehopqa} & \textbf{Average} \\
    \hline
    \rowcolor{gray!25}
    \multicolumn{3}{l}{\textit{Llama3}} & & & \\
    Llama-3.2-1B-Instruct & 26.27 & 37.29 & 17.99 & 25.31 & 26.72 \\
    Llama-3.2-3B-Instruct & 39.26 & 46.86 & 25.53 & 50.98 & 40.66 \\
    Llama-3.1-8B-Instruct & 56.73 & 77.90 & 42.83 & 56.89 & 58.59 \\
    Llama-3.3-70B-Instruct & 68.71 & 86.56 & 56.09 & 74.96 & 71.58 \\
    \hdashline
    \rowcolor{gray!25}
    \multicolumn{3}{l}{\textit{Qwen3}} & & & \\
    Qwen3-0.6B & 10.45 & 1.29 & 2.42 & 0.27 & 3.61 \\
    Qwen3-1.7B & 40.50 & 59.31 & 34.10 & 51.57 & 46.37 \\
    Qwen3-4B & 54.59 & 79.82 & 42.17 & 62.34 & 59.73 \\
    Qwen3-8B & 57.44 & 82.82 & 43.21 & 66.55 & 62.51 \\
    Qwen3-14B & 57.82 & 82.67 & 44.85 & 66.73 & 63.02 \\
    Qwen3-32B & 60.24 & 83.15 & 47.65 & 69.59 & 65.16 \\
    \bottomrule
  \end{tabular}
  \caption{Retrieval recall performance of off-the-shelf LLMs employing our designed hybrid decoding strategy via few-shot prompting in a training-free setting.}
  \label{tab:zeroshot}
\end{table*}

\subsection{Efficiency Comparison}
\label{sec:efficiency}

We compare inference efficiency by measuring average latency per query, with results visualized in Figure~\ref{fig:speed}.
ThinkGR exhibits higher efficiency than LLM-Driven Multi-Step Retrieval methods, as its end-to-end generation design eliminates the latency incurred by sequential LLM-retriever invocations.
While implicit methods like MDR achieve lower latency by avoiding explicit token generation, ThinkGR strikes a compelling balance: generating explicit thought tokens incurs only moderate latency cost while substantially improving retrieval performance on complex queries.
Additionally, the FM-index used in ThinkGR is highly space-efficient, requiring significantly less storage than dense retrieval index (detailed in Appendix~\ref{appendix:memory}).

\begin{figure}[t]
    \centering
    \includegraphics[width=\columnwidth]{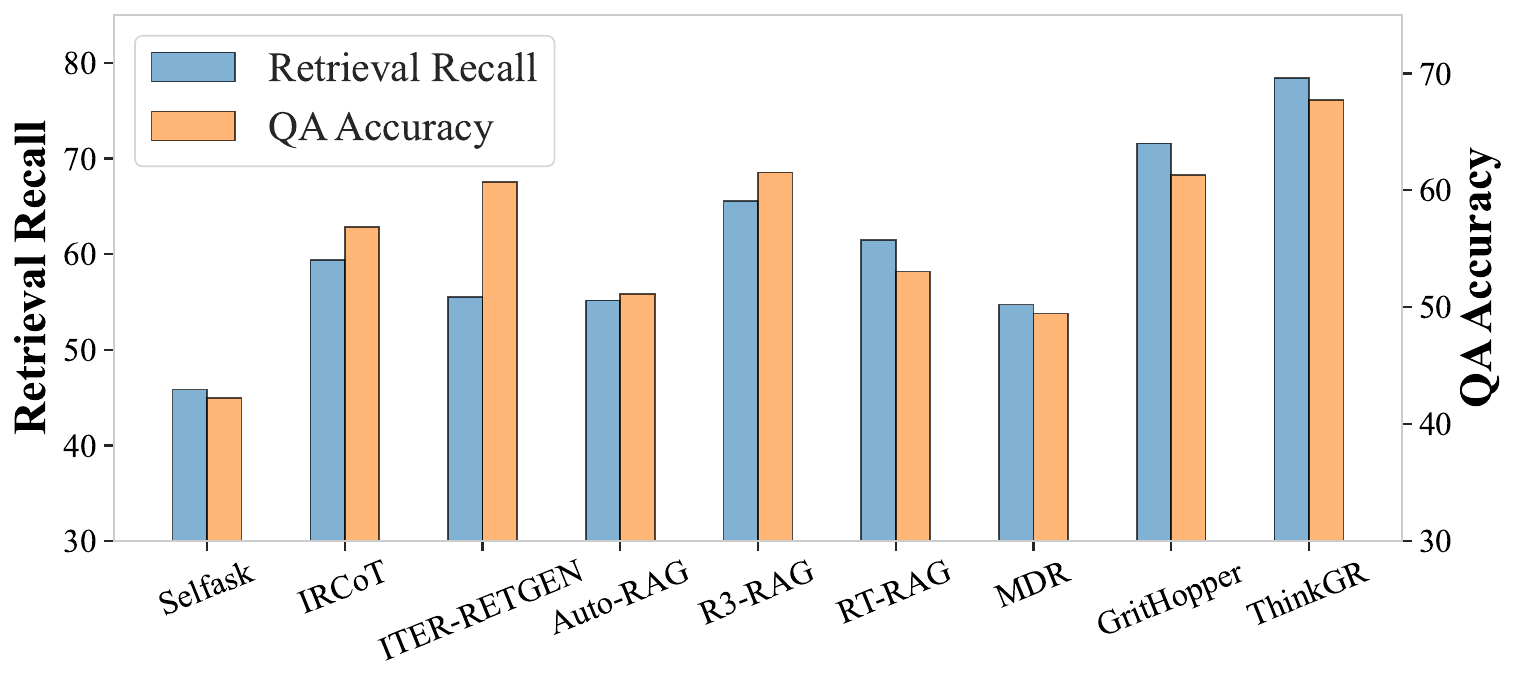}
    \caption{Comprehensive evaluation of retrieval and QA performance.
    The blue bars show average retrieval recall, while the orange bars represent QA accuracy.
    }
    \label{fig:qa}
\end{figure}

\subsection{Impact on Downstream QA}
\label{sec:qa}

To comprehensively evaluate the effectiveness of our method, we further assess ThinkGR's performance in the complete question answering task.
Following established practice in RAG-based QA evaluation, we employ Llama-3.3-70B-Instruct to generate answers based on the retrieved documents and the original question.
We use accuracy (Acc) as QA metric, which is determined by further prompting the LLM for judgment.
As visually summarized in Figure~\ref{fig:qa}, ThinkGR achieves the highest average QA accuracy across four benchmarks.
Its average accuracy exceeds the strongest baseline by 6.37\%, indicating that the quality of retrieved documents directly determines the accuracy of downstream answers.
Notably, the QA accuracy does not strictly correlate with the retrieval recall.
For example, IRCoT achieves a higher retrieval recall than ITER-RETGEN, but the latter outperforms in QA accuracy.
This discrepancy arises from the varying number of documents retrieved by different methods.
Retrieving more documents can increase recall, but it may also introduce irrelevant documents that negatively impact QA accuracy.
Overall, the experimental results show that ThinkGR effectively enhances retrieval quality without compromising downstream QA performance.
This further demonstrates ThinkGR's effectiveness as an end-to-end solution for complex information retrieval tasks.
The complete results are provided in Table~\ref{tab:qa}.

\subsection{Performance of Off-the-Shelf LLMs with Few-Shot Prompting}
\label{appendix:zeroshot}

To investigate the effectiveness of our proposed thought-augmented retrieval paradigm in a training-free setting, we examine whether off-the-shelf LLMs can perform hybrid decoding through prompt engineering.
We conduct experiments using carefully designed prompts and few-shot demonstrations.
Specifically, we test 10 LLMs from the Llama3 (1B - 70B) and Qwen3 (0.6B - 32B) families, providing the same instructions and context examples to guide them toward the desired output structure.
Since these off-the-shelf models have not been fine-tuned to generate our specific special tokens (e.g., \texttt{<docid\_start>}), we use square brackets \texttt{[} and \texttt{]} as substitutes for the start and end tokens of document identifiers.
The results in Table~\ref{tab:zeroshot} reveal three observations:
(1) Off-the-shelf LLMs exhibit reasonable performance (e.g., Llama3.3-70B achieves 71.58\%), validating the inherent feasibility of our thought-augmented retrieval paradigm.
This demonstrates the effectiveness of our designed paradigm even in a training-free setting.
(2) Model scale determines feasibility: while larger LLM show progressively better performance, smaller LLM like Qwen3-0.6B fail catastrophically due to insufficient instruction-following ability.
(3) Despite promising results, all these models underperform our trained ThinkGR (78.44\%), with even the strongest Llama3.3-70B trailing by 6.86\%.
This demonstrates the effectiveness of our designed two-phase training strategy.

\section{Conclusion}

This paper presents a preliminary exploration of integrating chain-of-thought into generative retrieval.
While existing approaches directly map queries to document identifiers, we demonstrate that interleaving explicit thought processes with retrieval actions can substantially improve performance on complex queries.
To realize this idea, we develop ThinkGR as a preliminary instantiation, incorporating a hybrid decoding strategy to unify free-form thought with constrained identifier generation, and a two-phase training approach that leverages retrieval-grounded reinforcement learning to optimize thought quality.
Empirical results across four multi-hop benchmarks validate our approach, with ThinkGR achieving an average improvement of +6.86\% over state-of-the-art baselines.
We hope this preliminary study demonstrates the feasibility and potential of thought-augmented generative retrieval, inspiring further investigation into deliberative generation for information retrieval systems.

\section*{Limitations}

As an early exploration into integrating chain-of-thought with generative retrieval, our study has several limitations that suggest directions for future research:
\begin{inparaenum}[(1)]
\item Our current implementation relies on established training algorithms (SFT and KTO) to align thought and retrieval.
While effective for this preliminary exploration, these methods may not fully unlock the potential of thought-augmented retrieval.
Future work could investigate more specialized optimization techniques, such as process reward models that provide fine-grained feedback on thought quality, or curriculum learning strategies that progressively increase thought complexity.
\item Although our method achieves state-of-the-art results on multi-hop retrieval, it is specifically tailored for these scenarios through its triple-based docid design.
This specialization may lead to suboptimal performance on standard one-hop benchmarks (e.g., Natural Questions) where intermediate reasoning can introduce unnecessary complexity.
Moreover, the current triple representation primarily captures entity-relation structures, potentially overlooking non-factual content.
Future work should explore developing more generalizable docid representations (e.g., entity-fact tuples) that can effectively capture both relational semantics for multi-hop traversal and broader content for one-hop matching.
\item To rigorously evaluate our method's effectiveness in complex reasoning scenarios, our experiments primarily compare against approaches specifically designed for multi-hop retrieval, which represent stronger baselines in this setting.
While this comparison strategy validates ThinkGR's advantages in its target domain, a comprehensive evaluation of thought-augmented generative retrieval as a general paradigm requires broader analysis.
In future work, we plan to conduct more detailed comparisons with existing generative retrieval methods across diverse retrieval tasks to better characterize the benefits and trade-offs of integrating chain-of-thought into the generative retrieval framework.
\end{inparaenum}

\section*{Ethics Statement}

This work explores the integration of Chain-of-Thought reasoning into generative retrieval models to enhance their retrieval capabilities, particularly for handling complex queries.
All experiments derived in this study rely on publicly available datasets and do not involve any collection of personal data or interaction with human subjects.



\bibliography{main}

@inproceedings{yang-etal-2018-hotpotqa,
    title = "{H}otpot{QA}: A Dataset for Diverse, Explainable Multi-hop Question Answering",
    author = "Yang, Zhilin  and
      Qi, Peng  and
      Zhang, Saizheng  and
      Bengio, Yoshua  and
      Cohen, William  and
      Salakhutdinov, Ruslan  and
      Manning, Christopher D.",
    editor = "Riloff, Ellen  and
      Chiang, David  and
      Hockenmaier, Julia  and
      Tsujii, Jun{'}ichi",
    booktitle = "Proceedings of the 2018 Conference on Empirical Methods in Natural Language Processing",
    month = oct # "-" # nov,
    year = "2018",
    address = "Brussels, Belgium",
    publisher = "Association for Computational Linguistics",
    url = "https://aclanthology.org/D18-1259/",
    doi = "10.18653/v1/D18-1259",
    pages = "2369--2380"
}

@article{trivedi2022musique,
  title={MuSiQue: Multihop Questions via Single-hop Question Composition},
  author={Trivedi, Harsh and Balasubramanian, Niranjan and Khot, Tushar and Sabharwal, Ashish},
  journal={Transactions of the Association for Computational Linguistics},
  volume={10},
  pages={539--554},
  year={2022},
  publisher={MIT Press One Broadway, 12th Floor, Cambridge, Massachusetts 02142, USA~…}
}

@inproceedings{trivedi-etal-2023-interleaving,
    title = "Interleaving Retrieval with Chain-of-Thought Reasoning for Knowledge-Intensive Multi-Step Questions",
    author = "Trivedi, Harsh  and
      Balasubramanian, Niranjan  and
      Khot, Tushar  and
      Sabharwal, Ashish",
    editor = "Rogers, Anna  and
      Boyd-Graber, Jordan  and
      Okazaki, Naoaki",
    booktitle = "Proceedings of the 61st Annual Meeting of the Association for Computational Linguistics (Volume 1: Long Papers)",
    month = jul,
    year = "2023",
    address = "Toronto, Canada",
    publisher = "Association for Computational Linguistics",
    url = "https://aclanthology.org/2023.acl-long.557/",
    doi = "10.18653/v1/2023.acl-long.557",
    pages = "10014--10037"
}

@article{li2025r3,
  title={R3-RAG: Learning Step-by-Step Reasoning and Retrieval for LLMs via Reinforcement Learning},
  author={Li, Yuan and Luo, Qi and Li, Xiaonan and Li, Bufan and Cheng, Qinyuan and Wang, Bo and Zheng, Yining and Wang, Yuxin and Yin, Zhangyue and Qiu, Xipeng},
  journal={arXiv preprint arXiv:2505.23794},
  year={2025}
}

@article{xiong2020answering,
  title={Answering Complex Open-Domain Questions with Multi-Hop Dense Retrieval},
  author={Xiong, Wenhan and Li, Xiang Lorraine and Iyer, Srinivasan and Du, Jingfei and Lewis, Patrick and Wang, William Yang and Mehdad, Yashar and Yih, Wen-tau and Riedel, Sebastian and Kiela, Douwe and O{\u{g}}uz, Barlas},
  journal={International Conference on Learning Representations},
  year={2021}
}

@article{erker2025grithopper,
  title={GRITHopper: Decomposition-Free Multi-Hop Dense Retrieval},
  author={Erker, Justus-Jonas and Reimers, Nils and Gurevych, Iryna},
  journal={arXiv preprint arXiv:2503.07519},
  year={2025}
}

@article{jaech2024openai,
  title={Openai o1 system card},
  author={Jaech, Aaron and Kalai, Adam and Lerer, Adam and Richardson, Adam and El-Kishky, Ahmed and Low, Aiden and Helyar, Alec and Madry, Aleksander and Beutel, Alex and Carney, Alex and others},
  journal={arXiv preprint arXiv:2412.16720},
  year={2024}
}

@article{guo2025deepseek,
  title={Deepseek-r1: Incentivizing reasoning capability in llms via reinforcement learning},
  author={Guo, Daya and Yang, Dejian and Zhang, Haowei and Song, Junxiao and Zhang, Ruoyu and Xu, Runxin and Zhu, Qihao and Ma, Shirong and Wang, Peiyi and Bi, Xiao and others},
  journal={arXiv preprint arXiv:2501.12948},
  year={2025}
}

@article{wei2022chain,
  title={Chain-of-thought prompting elicits reasoning in large language models},
  author={Wei, Jason and Wang, Xuezhi and Schuurmans, Dale and Bosma, Maarten and Xia, Fei and Chi, Ed and Le, Quoc V and Zhou, Denny and others},
  journal={Advances in neural information processing systems},
  volume={35},
  pages={24824--24837},
  year={2022}
}

@article{grattafiori2024llama,
  title={The llama 3 herd of models},
  author={Grattafiori, Aaron and Dubey, Abhimanyu and Jauhri, Abhinav and Pandey, Abhinav and Kadian, Abhishek and Al-Dahle, Ahmad and Letman, Aiesha and Mathur, Akhil and Schelten, Alan and Vaughan, Alex and others},
  journal={arXiv preprint arXiv:2407.21783},
  year={2024}
}

@inproceedings{press-etal-2023-measuring,
    title = "Measuring and Narrowing the Compositionality Gap in Language Models",
    author = "Press, Ofir  and
      Zhang, Muru  and
      Min, Sewon  and
      Schmidt, Ludwig  and
      Smith, Noah  and
      Lewis, Mike",
    editor = "Bouamor, Houda  and
      Pino, Juan  and
      Bali, Kalika",
    booktitle = "Findings of the Association for Computational Linguistics: EMNLP 2023",
    month = dec,
    year = "2023",
    address = "Singapore",
    publisher = "Association for Computational Linguistics",
    url = "https://aclanthology.org/2023.findings-emnlp.378/",
    doi = "10.18653/v1/2023.findings-emnlp.378",
    pages = "5687--5711"
}

@inproceedings{yao2023react,
  title={React: Synergizing reasoning and acting in language models},
  author={Yao, Shunyu and Zhao, Jeffrey and Yu, Dian and Du, Nan and Shafran, Izhak and Narasimhan, Karthik and Cao, Yuan},
  booktitle={International Conference on Learning Representations (ICLR)},
  year={2023}
}

@inproceedings{shao-etal-2023-enhancing,
    title = "Enhancing Retrieval-Augmented Large Language Models with Iterative Retrieval-Generation Synergy",
    author = "Shao, Zhihong  and
      Gong, Yeyun  and
      Shen, Yelong  and
      Huang, Minlie  and
      Duan, Nan  and
      Chen, Weizhu",
    editor = "Bouamor, Houda  and
      Pino, Juan  and
      Bali, Kalika",
    booktitle = "Findings of the Association for Computational Linguistics: EMNLP 2023",
    month = dec,
    year = "2023",
    address = "Singapore",
    publisher = "Association for Computational Linguistics",
    url = "https://aclanthology.org/2023.findings-emnlp.620/",
    doi = "10.18653/v1/2023.findings-emnlp.620",
    pages = "9248--9274"
}

@article{yu2024autorag,
      title={Auto-RAG: Autonomous Retrieval-Augmented Generation for Large Language Models}, 
      author={Tian Yu and Shaolei Zhang and Yang Feng},
      year={2024},
      eprint={2411.19443},
      archivePrefix={arXiv},
      primaryClass={cs.CL},
      url={https://arxiv.org/abs/2411.19443}, 
}

@article{guan2025deeprag,
  title={DeepRAG: Thinking to Retrieval Step by Step for Large Language Models},
  author={Guan, Xinyan and Zeng, Jiali and Meng, Fandong and Xin, Chunlei and Lu, Yaojie and Lin, Hongyu and Han, Xianpei and Sun, Le and Zhou, Jie},
  journal={arXiv preprint arXiv:2502.01142},
  year={2025}
}

@article{yan2025o1,
  title={O1 embedder: Let retrievers think before action},
  author={Yan, Ruiran and Liu, Zheng and Lian, Defu},
  journal={arXiv preprint arXiv:2502.07555},
  year={2025}
}

@inproceedings{deautoregressive,
  title={Autoregressive Entity Retrieval},
  author={De Cao, Nicola and Izacard, Gautier and Riedel, Sebastian and Petroni, Fabio},
  booktitle={International Conference on Learning Representations},
  year={2021}
}

@article{bevilacqua2022autoregressive,
  title={Autoregressive search engines: Generating substrings as document identifiers},
  author={Bevilacqua, Michele and Ottaviano, Giuseppe and Lewis, Patrick and Yih, Scott and Riedel, Sebastian and Petroni, Fabio},
  journal={Advances in Neural Information Processing Systems},
  volume={35},
  pages={31668--31683},
  year={2022}
}

@article{zhou2022ultron,
  title={Ultron: An ultimate retriever on corpus with a model-based indexer},
  author={Zhou, Yujia and Yao, Jing and Dou, Zhicheng and Wu, Ledell and Zhang, Peitian and Wen, Ji-Rong},
  journal={arXiv preprint arXiv:2208.09257},
  year={2022}
}

@inproceedings{zhang2024generative,
  title={Generative retrieval via term set generation},
  author={Zhang, Peitian and Liu, Zheng and Zhou, Yujia and Dou, Zhicheng and Liu, Fangchao and Cao, Zhao},
  booktitle={Proceedings of the 47th International ACM SIGIR Conference on Research and Development in Information Retrieval},
  pages={458--468},
  year={2024}
}

@inproceedings{NEURIPS2023_91228b94,
  author = {Sun, Weiwei and Yan, Lingyong and Chen, Zheng and Wang, Shuaiqiang and Zhu, Haichao and Ren, Pengjie and Chen, Zhumin and Yin, Dawei and Rijke, Maarten and Ren, Zhaochun},
  booktitle = {Advances in Neural Information Processing Systems},
  editor = {A. Oh and T. Naumann and A. Globerson and K. Saenko and M. Hardt and S. Levine},
  pages = {46345--46361},
  publisher = {Curran Associates, Inc.},
  title = {Learning to Tokenize for Generative Retrieval},
  volume = {36},
  year = {2023}
}

@inproceedings{zhang2025excluir,
  title={Excluir: Exclusionary neural information retrieval},
  author={Zhang, Wenhao and Zhang, Mengqi and Wu, Shiguang and Pei, Jiahuan and Ren, Zhaochun and de Rijke, Maarten and Chen, Zhumin and Ren, Pengjie},
  booktitle={Proceedings of the AAAI Conference on Artificial Intelligence},
  volume={39},
  number={12},
  pages={13295--13303},
  year={2025}
}

@inproceedings{lee-etal-2022-generative,
    title = "Generative Multi-hop Retrieval",
    author = "Lee, Hyunji  and
      Yang, Sohee  and
      Oh, Hanseok  and
      Seo, Minjoon",
    editor = "Goldberg, Yoav  and
      Kozareva, Zornitsa  and
      Zhang, Yue",
    booktitle = "Proceedings of the 2022 Conference on Empirical Methods in Natural Language Processing",
    month = dec,
    year = "2022",
    address = "Abu Dhabi, United Arab Emirates",
    publisher = "Association for Computational Linguistics",
    url = "https://aclanthology.org/2022.emnlp-main.92/",
    doi = "10.18653/v1/2022.emnlp-main.92",
    pages = "1417--1436"
}

@article{tay2022transformer,
  title={Transformer memory as a differentiable search index},
  author={Tay, Yi and Tran, Vinh and Dehghani, Mostafa and Ni, Jianmo and Bahri, Dara and Mehta, Harsh and Qin, Zhen and Hui, Kai and Zhao, Zhe and Gupta, Jai and others},
  journal={Advances in Neural Information Processing Systems},
  volume={35},
  pages={21831--21843},
  year={2022}
}

@article{ethayarajh2024kto,
  title={Kto: Model alignment as prospect theoretic optimization},
  author={Ethayarajh, Kawin and Xu, Winnie and Muennighoff, Niklas and Jurafsky, Dan and Kiela, Douwe},
  journal={arXiv preprint arXiv:2402.01306},
  year={2024}
}

@inproceedings{ferragina2000opportunistic,
  title={Opportunistic data structures with applications},
  author={Ferragina, Paolo and Manzini, Giovanni},
  booktitle={Proceedings 41st annual symposium on foundations of computer science},
  pages={390--398},
  year={2000},
  organization={IEEE}
}

@inproceedings{ho-etal-2020-constructing,
    title = "Constructing A Multi-hop {QA} Dataset for Comprehensive Evaluation of Reasoning Steps",
    author = "Ho, Xanh  and
      Duong Nguyen, Anh-Khoa  and
      Sugawara, Saku  and
      Aizawa, Akiko",
    editor = "Scott, Donia  and
      Bel, Nuria  and
      Zong, Chengqing",
    booktitle = "Proceedings of the 28th International Conference on Computational Linguistics",
    month = dec,
    year = "2020",
    address = "Barcelona, Spain (Online)",
    publisher = "International Committee on Computational Linguistics",
    url = "https://aclanthology.org/2020.coling-main.580/",
    doi = "10.18653/v1/2020.coling-main.580",
    pages = "6609--6625"
}

@article{schnitzler2024morehopqa,
  title={Morehopqa: More than multi-hop reasoning},
  author={Schnitzler, Julian and Ho, Xanh and Huang, Jiahao and Boudin, Florian and Sugawara, Saku and Aizawa, Akiko},
  journal={arXiv preprint arXiv:2406.13397},
  year={2024}
}

@misc{izacard2021contriever,
  title={Unsupervised Dense Information Retrieval with Contrastive Learning}, 
  author={Gautier Izacard and Mathilde Caron and Lucas Hosseini and Sebastian Riedel and Piotr Bojanowski and Armand Joulin and Edouard Grave},
  year={2021},
  url = {https://arxiv.org/abs/2112.09118},
  doi = {10.48550/ARXIV.2112.09118},
}

@article{robertson2009probabilistic,
  title={The probabilistic relevance framework: BM25 and beyond},
  author={Robertson, Stephen and Zaragoza, Hugo and others},
  journal={Foundations and Trends{\textregistered} in Information Retrieval},
  volume={3},
  number={4},
  pages={333--389},
  year={2009},
  publisher={Now Publishers, Inc.}
}

@misc{bge_embedding,
  title={C-Pack: Packaged Resources To Advance General Chinese Embedding}, 
  author={Shitao Xiao and Zheng Liu and Peitian Zhang and Niklas Muennighoff},
  year={2023},
  eprint={2309.07597},
  archivePrefix={arXiv},
  primaryClass={cs.CL}
}

@inproceedings{zheng2024llamafactory,
  title={LlamaFactory: Unified Efficient Fine-Tuning of 100+ Language Models},
  author={Yaowei Zheng and Richong Zhang and Junhao Zhang and Yanhan Ye and Zheyan Luo and Zhangchi Feng and Yongqiang Ma},
  booktitle={Proceedings of the 62nd Annual Meeting of the Association for Computational Linguistics (Volume 3: System Demonstrations)},
  address={Bangkok, Thailand},
  publisher={Association for Computational Linguistics},
  year={2024},
  url={http://arxiv.org/abs/2403.13372}
}

@article{FlashRAG,
  author = {Jiajie Jin and
      Yutao Zhu and
      Xinyu Yang and
      Chenghao Zhang and
      Zhicheng Dou},
  title  = {FlashRAG: {A} Modular Toolkit for Efficient Retrieval-Augmented Generation
      Research},
  journal      = {CoRR},
  volume = {abs/2405.13576},
  year   = {2024},
  url    = {https://doi.org/10.48550/arXiv.2405.13576},
  doi    = {10.48550/ARXIV.2405.13576},
  eprinttype    = {arXiv},
  eprint = {2405.13576},
  bibsource    = {dblp computer science bibliography, https://dblp.org}
}

@article{rafailov2023direct,
  title={Direct preference optimization: Your language model is secretly a reward model},
  author={Rafailov, Rafael and Sharma, Archit and Mitchell, Eric and Manning, Christopher D and Ermon, Stefano and Finn, Chelsea},
  journal={Advances in neural information processing systems},
  volume={36},
  pages={53728--53741},
  year={2023}
}

@article{schulman2017proximal,
  title={Proximal policy optimization algorithms},
  author={Schulman, John and Wolski, Filip and Dhariwal, Prafulla and Radford, Alec and Klimov, Oleg},
  journal={arXiv preprint arXiv:1707.06347},
  year={2017}
}

@article{zhang2025gem,
  title={GEM: Empowering LLM for both Embedding Generation and Language Understanding},
  author={Zhang, Caojin and Zhang, Qiang and Li, Ke and Nuthalapati, Sai Vidyaranya and Zhang, Benyu and Liu, Jason and Li, Serena and Zhang, Lizhu and Fan, Xiangjun},
  journal={arXiv preprint arXiv:2506.04344},
  year={2025}
}

@article{tang2025large,
  title={Large Reasoning Embedding Models: Towards Next-Generation Dense Retrieval Paradigm},
  author={Tang, Jianting and Li, Dongshuai and Wen, Tao and Lv, Fuyu and Ou, Dan and Xu, Linli},
  journal={arXiv preprint arXiv:2510.14321},
  year={2025}
}

@article{zhang2025does,
  title={Does Generative Retrieval Overcome the Limitations of Dense Retrieval?},
  author={Zhang, Yingchen and Zhang, Ruqing and Guo, Jiafeng and de Rijke, Maarten and Fan, Yixing and Cheng, Xueqi},
  journal={arXiv preprint arXiv:2509.22116},
  year={2025}
}

@article{kwiatkowski2019natural,
  title={Natural questions: a benchmark for question answering research},
  author={Kwiatkowski, Tom and Palomaki, Jennimaria and Redfield, Olivia and Collins, Michael and Parikh, Ankur and Alberti, Chris and Epstein, Danielle and Polosukhin, Illia and Devlin, Jacob and Lee, Kenton and others},
  journal={Transactions of the Association for Computational Linguistics},
  volume={7},
  pages={453--466},
  year={2019},
  publisher={MIT Press One Rogers Street, Cambridge, MA 02142-1209, USA journals-info~…}
}

@inproceedings{li2024corpuslm,
  title={Corpuslm: Towards a unified language model on corpus for knowledge-intensive tasks},
  author={Li, Xiaoxi and Dou, Zhicheng and Zhou, Yujia and Liu, Fangchao},
  booktitle={Proceedings of the 47th International ACM SIGIR Conference on Research and Development in Information Retrieval},
  pages={26--37},
  year={2024}
}

@inproceedings{dong2026multi,
  title={Multi-Step Semantic Reasoning in Generative Retrieval},
  author={Dong, Steven and Tang, Yubao and de Rijke, Maarten},
  booktitle={European Conference on Information Retrieval},
  pages={273--281},
  year={2026},
  organization={Springer}
}

@article{liu2025onerec,
  title={Onerec-think: In-text reasoning for generative recommendation},
  author={Liu, Zhanyu and Wang, Shiyao and Wang, Xingmei and Zhang, Rongzhou and Deng, Jiaxin and Bao, Honghui and Zhang, Jinghao and Li, Wuchao and Zheng, Pengfei and Wu, Xiangyu and others},
  journal={arXiv preprint arXiv:2510.11639},
  year={2025}
}

@article{zhang2025retrieval,
  title={Retrieval-in-the-Chain: Bootstrapping Large Language Models for Generative Retrieval},
  author={Zhang, Yingchen and Zhang, Ruqing and Guo, Jiafeng and Peng, Wenjun and Li, Sen and Lv, Fuyu},
  journal={arXiv preprint arXiv:2510.13095},
  year={2025}
}

@article{shi2026reasoning,
  title={Reasoning in Trees: Improving Retrieval-Augmented Generation for Multi-Hop Question Answering},
  author={Shi, Yuling and Sun, Maolin and Liu, Zijun and Yang, Mo and Fang, Yixiong and Sun, Tianran and Gu, Xiaodong},
  journal={arXiv preprint arXiv:2601.11255},
  year={2026}
}

\clearpage
\appendix

\section{Prompt Templates}
\label{appendix:prompt}
\begin{figure*}
    \centering
    \includegraphics[width=2\columnwidth]{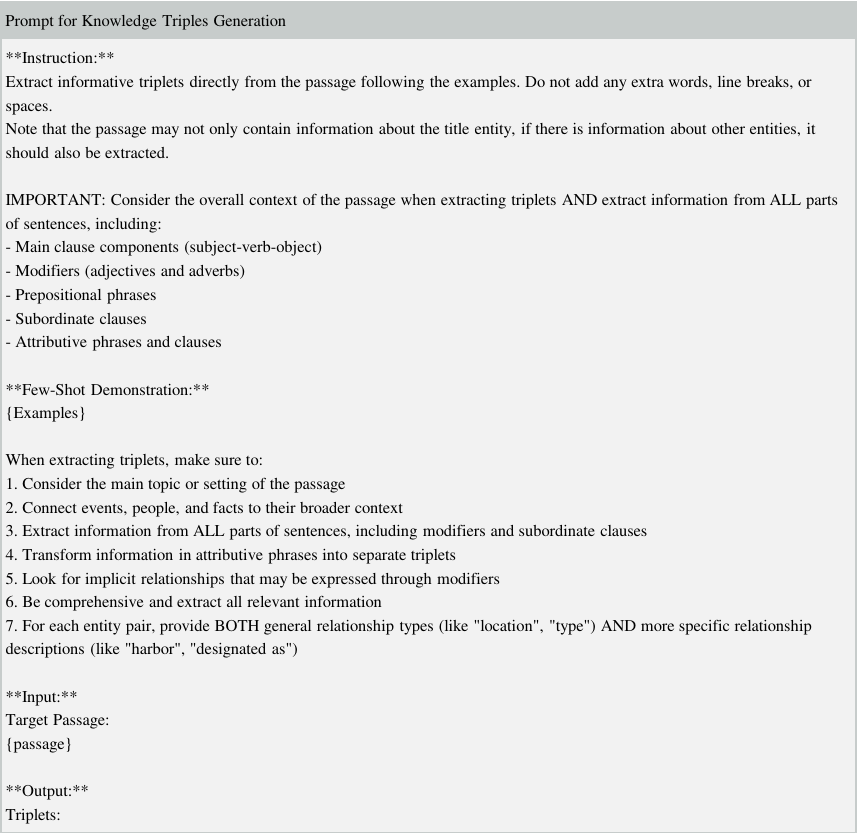}
    \caption{Prompt templates for knowledge triples generation.}
    \label{fig:prompt_triples}
\end{figure*}

\begin{figure*}
    \centering
    \includegraphics[width=2\columnwidth]{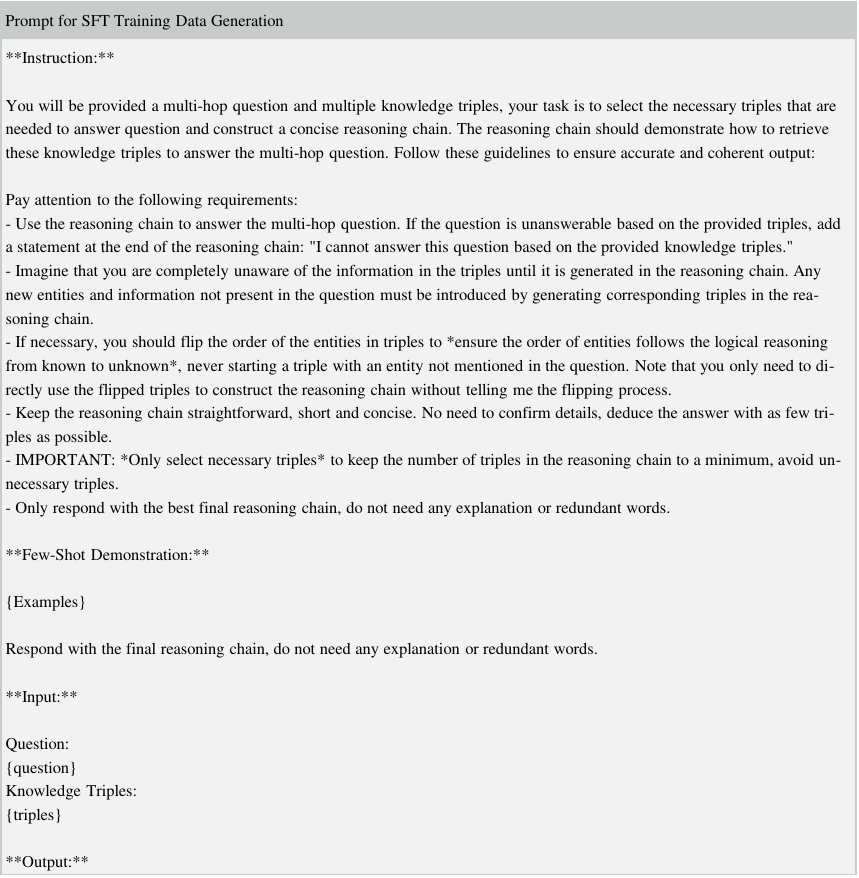}
    \caption{Prompt templates for SFT training data generation.}
    \label{fig:prompt_sft}
\end{figure*}

In this section, we present the prompt used to instruct LLM in summarizing documents into multiple knowledge triples (Figure~\ref{fig:prompt_triples}), and the prompt used to guide LLM in generating thought-retrieval chains for SFT training data (Figure~\ref{fig:prompt_sft}).

\section{Detailed Statistics of Datasets}
\label{appendix:datasets}

Among the datasets used, HotpotQA and MoreHopQA involve 2-hop questions, MuSiQue covers questions with 2-4 hops, while 2WikiMultiHopQA has up to 6 ground-truth documents for one question.
MoreHopQA does not provide a training set, thus it is used to evaluate the model's performance in out-of-domain setting.
It is worth noting that HotpotQA has been found to have the over-specification issue~\cite{trivedi2022musique}, where the questions contain too many ground-truth document fragments, leading to high lexical overlap between questions and supporting documents.
This allows models to easily identify relevant documents by matching fragments rather than performing multi-step thought, making it difficult to accurately assess their capabilities.
We show some examples of HotpotQA in Appendix~\ref{appendix:over_specification} to illustrate this issue and further discuss its impact in the experimental results analysis.
Nonetheless, we still include it as a reference for evaluation.

To construct corpus, for each dataset, we collected the supporting and non-supporting documents provided for each instance and used them as the corpus.
Since the original corpus of MoreHopQA contains only about 1K documents, which deviates from real-world retrieval scenarios, we combine it with the 2WikiMultiHopQA corpus to increase the difficulty and ensure the reliability of the evaluation.

Table~\ref{tab:dataset} provides detailed statistics of the datasets used in our experiments, including corpus size, number of training/test data, and the number of hops covered by the queries.

\begin{table}[t]
    \centering
    \small
    \begin{tabular}{lcccc}
    \toprule
    Dataset & HotpotQA & 2Wiki. & Musique & Morehopqa \\
    \midrule
    \# Corpus & 5M & 398K & 118K & 399K \\
    \# Train & 84,812 & 167,181 & 19,116 & - \\
    \# Test & 5,447 & 12,576 & 2,417 & 1,118 \\
    \# Hops & 2 & 2-6 & 2-4 & 2 \\
    \bottomrule
    \end{tabular}
    \caption{Detailed statistics of the datasets used in our experiments, including corpus size, number of training/test data, and the number of hops covered by the queries.}
    \label{tab:dataset}
\end{table}

\section{Implementation Details}
\label{appendix:implementation}

ThinkGR is trained using Llama-3.1-8B-Instruct~\cite{grattafiori2024llama} as the base model.
We employ the Llama-Factory framework~\cite{zheng2024llamafactory} for sft training, using DeepSpeed zero-3 for full fine-tuning with a learning rate of 1e-6, batch size of 512, cutoff length of 2048, and warmup ratio of 0.05.
For the \ac{RL} phase, we train ThinkGR using KTO with a learning rate of 4e-7, batch size of 128, and set the risk aversion hyperparameter $\beta$ to 0.1.
We select the HotpotQA and Musique datasets for training in this phase, as the model already achieves high retrieval accuracy on the 2WikiMultihopQA dataset after the first training phase.
The threshold $\tau$ for distinguishing desirable and undesirable responses is set to 0.5.
This threshold was selected based on pilot experiments to ensure a clear and robust learning signal.

For constructing semantic triples as document identifiers, we use Llama-3.1-8B-Instruct to summarize each document into knowledge triples.
For generating thought-retrieval chains as SFT training data, we employ Llama-3.3-70B-Instruct to ensure high-quality thought trajectories.

We reproduced baselines based on official code and FlashRAG~\cite{FlashRAG}.
To ensure fairness and comparability, we constrained all baselines to retrieve no more than 10 documents per query.
Specifically, for conventional non-iterative retrievers (incapable of multi-hop reasoning), we set the number of retrieved documents to the number of ground-truth documents plus one.
For iterative retrieval methods, we limited the number of hops to 5 and the number of documents retrieved per step to 2.
LLM-driven multi-step retrieval methods require an off-the-shelf LLM and retriever.
Except for R3-RAG and Auto-RAG, which use their own fine-tuned LLM, we uniformly use Llama-3.3-70B-Instruct as LLM and Contriever-MSMARCO~\cite{izacard2021contriever} as retriever to ensure fair comparison.
Since RT-RAG involves substantially more LLM calls per query, performing separate think, retrieve, verify, and reflect steps in a tree-structured pipeline, we reproduced it using Llama-3.1-8B-Instruct to maintain practical evaluation efficiency; this is consistent with the method's inference procedure, which interleaves multiple structured reasoning and retrieval operations.
All experiments are conducted on NVIDIA A800 GPUs.

\begin{table*}[ht]
  \centering
  \begin{tabular}{lccccc}
    \toprule
    \textbf{Base Model} & \textbf{HotpotQA} & \textbf{2WikiMultihopQA} & \textbf{Musique} & \textbf{Morehopqa} & \textbf{Average} \\
    \hline
    \rowcolor{gray!25}
    \multicolumn{3}{l}{\textit{Llama3}} & & & \\
    Llama3.1-8B & 76.09 & 93.19 & 63.98 & 80.50 & 78.44 \\
    Llama3.2-3B & 66.55 & 89.66 & 50.70 & 74.73 & 70.41 \\
    Llama3.2-1B & 60.94 & 88.43 & 45.88 & 71.47 & 66.68 \\
    \hdashline
    \rowcolor{gray!25}
    \multicolumn{3}{l}{\textit{Qwen3}} & & & \\
    Qwen3-14B & 70.43 & 91.05 & 53.98 & 76.12 & 72.90 \\
    Qwen3-8B & 66.70 & 89.90 & 51.54 & 75.13 & 70.82 \\
    Qwen3-4B & 65.90 & 89.01 & 50.50 & 72.81 & 69.56 \\
    Qwen3-1.7B & 62.48 & 88.46 & 46.99 & 72.50 & 67.61 \\
    Qwen3-0.6B & 60.14 & 87.22 & 45.74 & 71.06 & 66.04 \\
    \bottomrule
  \end{tabular}
  \caption{Experimental results for different base models.}
  \label{tab:base_model}
\end{table*}
\begin{table*}[ht]
    \centering
    \begin{tabular}{lcccc}
    \toprule
    \textbf{Method} & \textbf{HotpotQA} & \textbf{2WikiMultihopQA} & \textbf{Musique} & \textbf{Average} \\
    \hline
    \rowcolor{gray!25}
    \multicolumn{3}{l}{\textit{LLM-Driven Multi-Step Retrieval}} & & \\
    Selfask & 37.81 & 33.88 & 28.90 & 33.53 \\
    IRCoT & 9.26 & 9.66 & 8.66 & 9.19 \\
    ITER-RETGEN  & 50.87 & 45.82 & 51.55 & 49.41 \\
    Auto-RAG & 4.37 & 6.74 & 5.22 & 5.44 \\
    R3-RAG & 0.93 & 1.36 & 1.21 & 1.17 \\
    RT-RAG & 52.53 & 39.89 & 85.72 & 59.38 \\
    \hdashline
    \rowcolor{gray!25}
    \multicolumn{3}{l}{\textit{Reasoning-Augmented Dense Retrieval}} & & \\
    MDR & 0.13 & 0.06 & 0.05 & 0.08 \\
    GritHopper & 65.51 & 3.53 & 1.36 & 23.47 \\
    \hdashline
    \rowcolor{gray!25}
    \multicolumn{3}{l}{\textit{Our Method}} & & \\
    ThinkGR & 1.13 & 0.60 & 1.38 & 1.04 \\
    \bottomrule
    \end{tabular}
    \caption{Experimental results of latency comparison.
    Lower values indicate superior efficiency.}
    \label{tab:speed}
\end{table*}

\begin{figure}[t]
    \centering
    \includegraphics[width=\columnwidth]{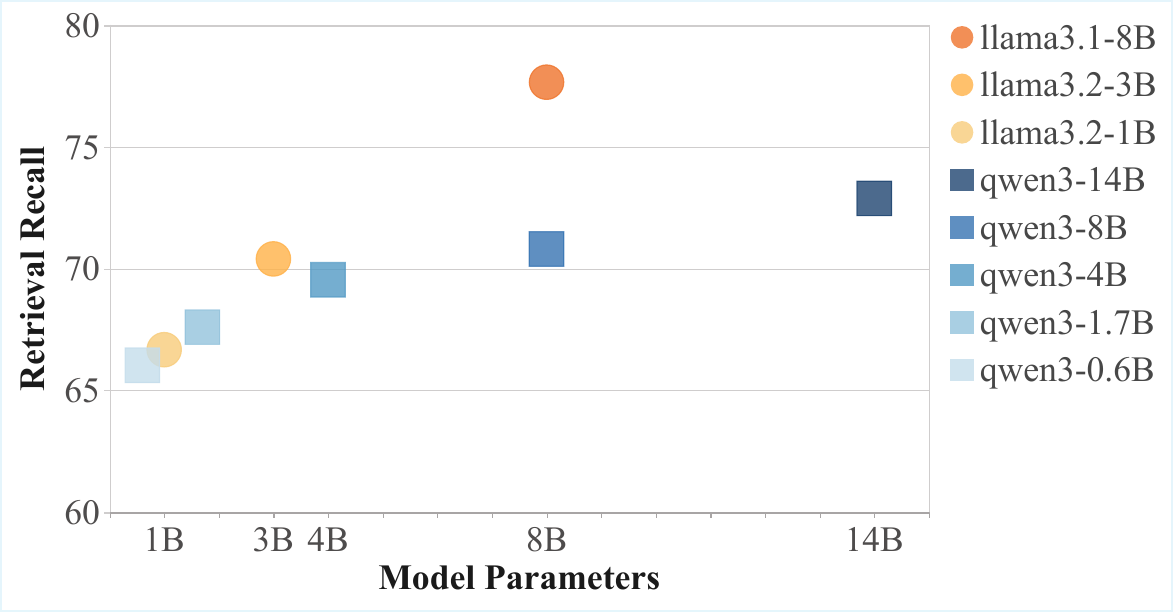}
    \caption{Performance of ThinkGR based on different base models.
    The x-axis represents the model size, while the y-axis shows the average retrieval recall across four datasets.}
    \label{fig:size}
\end{figure}

\section{Influence of Base Models}
\label{appendix:model}

\begin{figure}[t]
    \centering
    \includegraphics[width=\columnwidth]{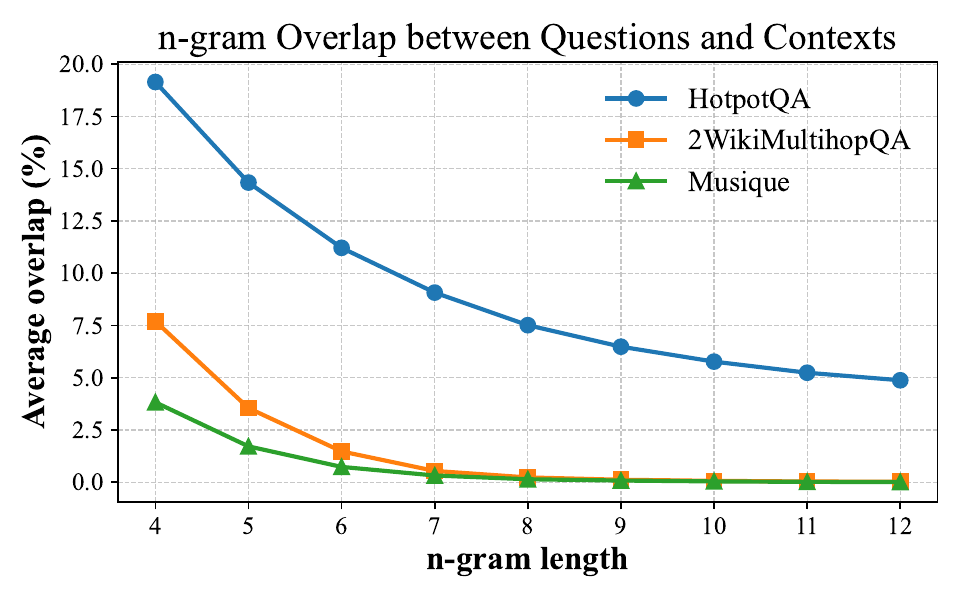}
    \caption{N-gram overlap between questions and ground-truth documents across three datasets. HotpotQA shows significantly higher overlap rates, indicating severe over-specification.}
    \label{fig:ngram_overlap}
\end{figure}

\begin{figure*}
    \centering
    \includegraphics[width=1.8\columnwidth]{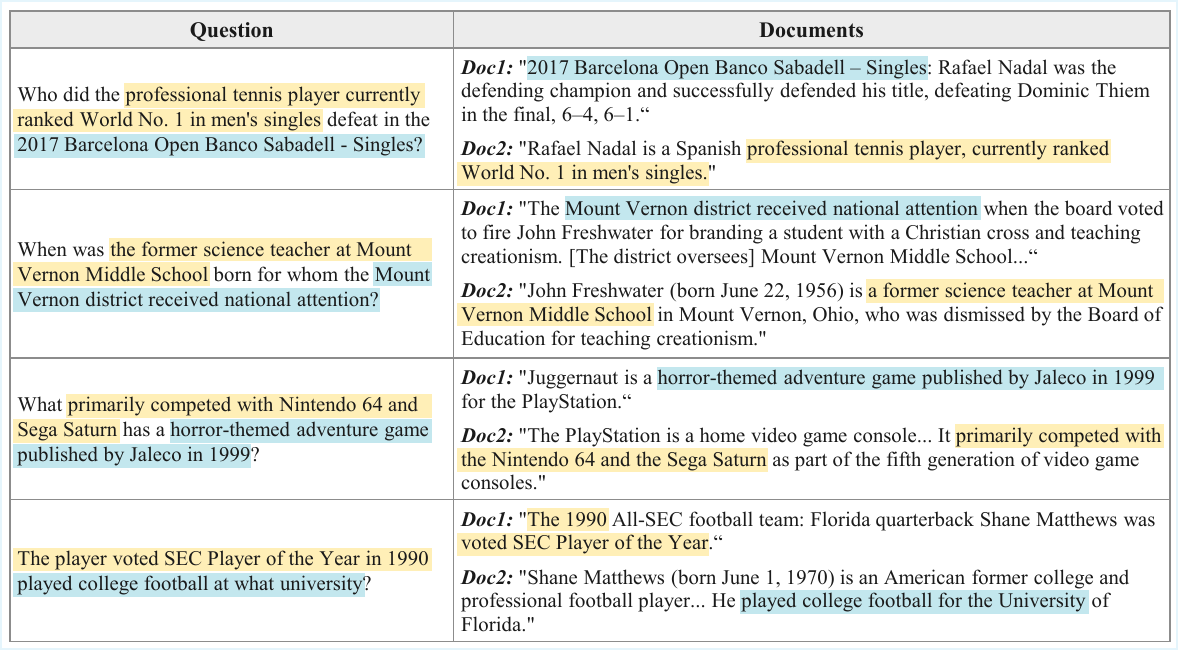}
    \caption{Examples of the over-specification issue in HotpotQA, where questions and supporting documents have high lexical overlap enabling direct matching without deep reasoning.}
    \label{fig:over_specification}
\end{figure*}

To assess the robustness of ThinkGR across different parameter scales, we systematically evaluate its performance using various base models.
We extend the Llama3.1-8B-Instruct model used in our main experiments to multiple models from the Llama3 and Qwen3 families, covering parameter scales ranging from 0.6B to 14B.
All models are trained under the same experimental settings, following our proposed two-phase training strategy.
The results presented in Figure~\ref{fig:size} demonstrate two key insights.
First, our method maintains satisfactory performance across different base models, even with smaller models like Qwen3-0.6B.
This indicates the architectural generality of our framework.
Second, under the same model architecture, larger models tend to perform better.
This is attributed to their stronger capabilities and richer parameter knowledge, which better support the complex thought-retrieval interactions.
These findings suggest that ThinkGR exhibits broad transferability across various base models.
While larger models yield better results, the framework remains practically feasible with smaller models, providing flexibility for deployment under different resource constraints.
Table~\ref{tab:base_model} shows the detailed performance of our method based on different base models.

\section{Complete Results of Efficiency Comparison}
\label{appendix:speed}

This section provides the complete experimental results of inference latency (seconds per query) across four datasets.
As detailed in Table~\ref{tab:speed}, ThinkGR demonstrates significantly reduced latency compared to LLM-Driven Multi-Step Retrieval methods, by eliminating costly sequential LLM/retriever calls.
While Reasoning-Augmented Dense Retrieval baselines generally show lower latency (0.08s average for MDR) due to implicit thought, ThinkGR achieves an optimal effectiveness-efficiency trade-off despite explicit thought generation.
Notably, GritHopper exhibits anomalously high latency on HotpotQA stems from hardware limitations: its GPU-accelerated FAISS retrieval defaults to CPU execution due to HotpotQA's index being too large for available GPU memory.
Crucially, in our method, the hybrid decoding strategy based on FM-index ensures constant time complexity, with latency determined solely by output token length.
This architecture fundamentally decouples inference speed from corpus scale, which is a critical advantage for real-world deployments with growing corpora.

\begin{table*}[t]
  \centering
  \begin{tabular}{lcccc}
    \toprule
    \textbf{Method} & \textbf{HotpotQA} & \textbf{2WikiMultihopQA} & \textbf{Musique} & \textbf{Morehopqa} \\
    \hline
    \rowcolor{gray!25}
    \multicolumn{4}{l}{\textit{LLM-Driven Multi-Step Retrieval}} & \\
    Selfask & 62.97 & 32.82 & 32.82 & 40.43 \\
    IRCoT & 74.78 & 58.08 & 47.75 & 46.78 \\
    ITER-RETGEN & 76.98 & 62.20 & 51.14 & \underline{52.42} \\
    Auto-RAG & 69.91 & 54.57 & 38.39 & 41.68 \\
    R3-RAG & 77.09 & \underline{70.59} & 47.12 & 51.34 \\
    RT-RAG & 70.48 & 56.46 & 40.86 & 44.44 \\
    \hdashline
    \rowcolor{gray!25}
    \multicolumn{4}{l}{\textit{Reasoning-Augmented Dense Retrieval}} & \\
    MDR & \underline{87.28} & 43.73 & 27.02 & 39.80 \\
    GritHopper & \textbf{91.08} & 53.10 & \underline{52.50} & 48.66 \\
    \hdashline
    \rowcolor{gray!25}
    \multicolumn{4}{l}{\textit{Our Method}} & \\
    ThinkGR & 79.99 & \textbf{79.52} & \textbf{57.94} & \textbf{53.40} \\
    \bottomrule
  \end{tabular}
  \caption{Experimental results for the complete QA task.
  We evaluate the QA performance using accuracy (Acc) as the metric, where Llama-3.3-70B-Instruct is used to answer questions based on the retrieved documents and judge the correctness.}
  \label{tab:qa}
\end{table*}
\begin{table*}[ht]
  \centering
  \begin{tabular}{lcccc}
    \toprule
    \textbf{Method} & \textbf{HotpotQA} & \textbf{2WikiMultihopQA} & \textbf{Musique} & \textbf{Morehopqa} \\
    \midrule
    R3-RAG (Contriever) & 58.56 & 82.82 & 51.70 & 70.44 \\
    R3-RAG (BGE-large) & 60.48 & 83.78 & 52.88 & 77.28 \\
    \textbf{ThinkGR} & \textbf{76.09} & \textbf{93.19} & \textbf{63.98} & \textbf{80.50} \\
    \bottomrule
  \end{tabular}
  \caption{Performance comparison of R3-RAG with different retrievers and ThinkGR.}
  \label{tab:retriever_comparison}
\end{table*}

\section{Analysis of Over-Specification Issue in HotpotQA}
\label{appendix:over_specification}

To quantitatively analyze the over-specification issue in HotpotQA, we conduct a statistical analysis of the n-gram overlap between questions and ground-truth documents across three datasets: HotpotQA, 2WikiMultihopQA, and Musique.
The results are presented in Figure~\ref{fig:ngram_overlap}.
As shown, HotpotQA exhibits significantly higher n-gram overlap rates compared to the other two datasets across all n-gram sizes (from 4-gram to 12-gram).
For instance, the average 4-gram overlap of HotpotQA reaches 19.15\%, which is 2.5$\times$ higher than 2WikiMultihopQA (7.70\%) and 5$\times$ higher than Musique (3.82\%).
This gap becomes even more pronounced for larger n-grams: at 8-gram, HotpotQA maintains 7.51\% overlap while 2WikiMultihopQA and Musique drop to only 0.23\% and 0.14\%, respectively.
These statistics confirm that HotpotQA suffers from severe over-specification, where questions contain substantial fragments from the ground-truth documents.
This allows models to retrieve relevant documents through simple lexical matching rather than performing genuine multi-step thought.

To further illustrate this issue, we present specific examples in Figure~\ref{fig:over_specification}.
These examples demonstrate how questions in HotpotQA often share significant lexical overlap with the supporting documents, allowing retrieval models to succeed through simple pattern matching rather than complex thought.

\section{Analysis of Triple Collisions}
\label{appendix:collision}

Since we use knowledge triples as document identifiers, it is possible for the same triple to appear in multiple documents.
We allow such collisions and retrieve all documents containing the matched triple.
Table~\ref{tab:collision} presents the proportion of triple overlaps across the four datasets.
As shown, the collision rate is very low (less than 3\% for triples appearing in $\ge$ 2 documents), indicating that triples are generally discriminative enough to identify documents.

\begin{table}[ht]
    \centering
    \small
    \begin{tabular}{lcc}
    \toprule
    \textbf{Dataset} & \textbf{Triples in $\ge$2 Docs} & \textbf{Triples in $\ge$3 Docs} \\
    \midrule
    HotpotQA & 0.91\% & 0.21\% \\
    2Wiki. & 1.80\% & 0.43\% \\
    Musique & 2.51\% & 0.45\% \\
    MoreHopQA & 1.79\% & 0.43\% \\
    \bottomrule
    \end{tabular}
    \caption{Proportion of triple overlaps across datasets.}
    \label{tab:collision}
\end{table}

\section{Analysis of Information Loss from Triple Extraction}
\label{appendix:info_loss}

A potential concern with our triple-based \ac{docids} representation is whether extracting knowledge triples from documents leads to significant information loss that could impair retrieval quality.
To directly evaluate this, we replace original documents with their extracted triples and test BGE-large~\cite{bge_embedding} on all four datasets.

Table~\ref{tab:info_loss} shows that retrieval performance with triples is nearly identical to that with original documents across all datasets.
Notably, performance on MuSiQue and MoreHopQA marginally improves with triple representations (+2.19\% and +1.03\%, respectively), suggesting that the structured relational format of triples can reduce noise from irrelevant document content and benefit entity-centric multi-hop queries.
These results demonstrate that our triple extraction process preserves the essential document semantics for retrieval with negligible information loss.

\begin{table}[ht]
    \centering
    \small
    \begin{tabular}{lcc}
    \toprule
    \textbf{Dataset} & \textbf{w/ Original Docs} & \textbf{w/ Triples} \\
    \midrule
    HotpotQA  & 60.48 & 57.42 \\
    2Wiki. & 58.43 & 56.66 \\
    Musique   & 33.39 & 35.58 \\
    MoreHopQA & 47.58 & 48.61 \\
    \bottomrule
    \end{tabular}
    \caption{Retrieval recall (\%) using original documents versus extracted triples as the corpus for BGE-large, showing negligible information loss from triple extraction.}
    \label{tab:info_loss}
\end{table}

\begin{figure}[ht]
    \centering
    \includegraphics[width=0.8\columnwidth]{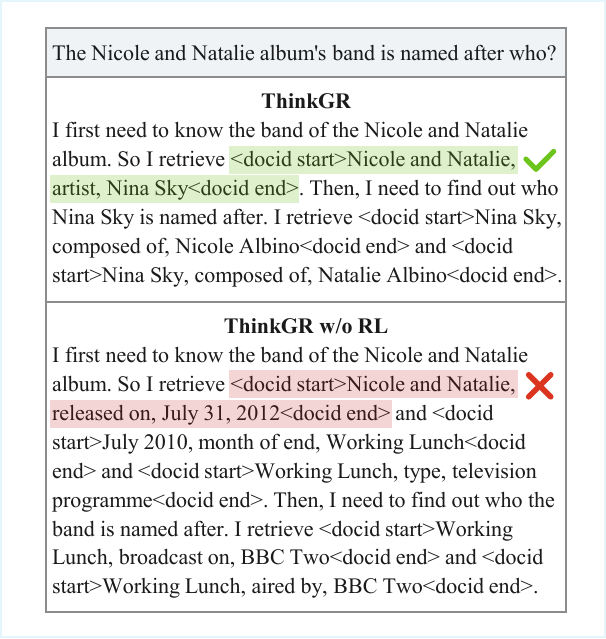}
    \caption{A case study on Musique illustrating the impact of retrieval-grounded thought optimization.
    ThinkGR (top) successfully generates the correct \ac{docids}, while the ablation variant without \ac{RL} (bottom) incorrectly predicts the first docid, leading to subsequent thought errors.
    }
    \label{fig:woRL}
\end{figure}

\section{Case Study Analysis}
\label{appendix:case_study}

We provide a specific case study in Figure~\ref{fig:woRL} to illustrate the impact of retrieval-grounded thought optimization.
ThinkGR (top) successfully generates the correct \ac{docids}, while the ablation variant without \ac{RL} (bottom) incorrectly predicts the first docid, leading to subsequent thought errors.

\section{Impact of Advanced Retriever on Baselines}
\label{appendix:retriever}

\begin{table*}[t]
    \centering
    \begin{tabular}{lccccc}
    \toprule
    \small
    \textbf{$\tau$} & \textbf{HotpotQA} & \textbf{2WikiMultihopQA} & \textbf{Musique} & \textbf{MoreHopQA} & \textbf{Average} \\
    \midrule
    0.1 & 74.05 & 92.82 & 61.93 & \textbf{80.72} & 77.38 \\
    0.3 & 74.67 & 92.61 & 62.56 & 79.70 & 77.39 \\
    0.5 & \textbf{76.09} & \textbf{93.19} & \textbf{63.98} & 80.50 & \textbf{78.44} \\
    0.7 & 74.89 & 93.00 & 63.24 & 80.23 & 77.84 \\
    \bottomrule
    \end{tabular}
    \caption{Sensitivity analysis of the threshold $\tau$ for distinguishing desirable and undesirable responses. Results are reported as Recall.}
    \label{tab:threshold}
\end{table*}

\begin{table*}[t]
    \centering
    \begin{tabular}{lcccc}
    \toprule
    \textbf{Dataset} & \textbf{Corpus Size} & \textbf{ThinkGR} & \textbf{GritHopper} & \textbf{Improvement} \\
    \midrule
    HotpotQA & 5M & 1.94 GB & 80 GB & \textbf{41x smaller} \\
    2WikiMultihopQA & 398K & 480 MB & 6.24 GB & \textbf{13x smaller} \\
    Musique & 118K & 68 MB & 1.85 GB & \textbf{27x smaller} \\
    MoreHopQA & 399K & 480 MB & 6.25 GB & \textbf{13x smaller} \\
    \bottomrule
    \end{tabular}
    \caption{Memory footprint comparison of the index between ThinkGR and GritHopper across different datasets.}
    \label{tab:memory}
\end{table*}

We investigate whether using a more advanced retriever can bridge the performance gap between LLM-Driven Multi-Step Retrieval methods and ThinkGR.
Specifically, we replace the Contriever in R3-RAG with BGE-large~\cite{bge_embedding} and evaluate its performance.
The results are presented in Table~\ref{tab:retriever_comparison}.
As shown, while the advanced retriever improves R3-RAG's performance, particularly on MorehopQA (+6.84\%), the gains on other datasets are limited.
ThinkGR still significantly outperforms R3-RAG equipped with BGE-large across all datasets.
This confirms that the superiority of ThinkGR stems from its unified thought-retrieval framework rather than the choice of the underlying retriever.

\section{Sensitivity Analysis of Threshold $\tau$}
\label{appendix:threshold}

The threshold $\tau$ controls the boundary for labeling model-generated responses as undesirable during the retrieval-grounded thought optimization phase.
To analyze the sensitivity of this hyperparameter, we conduct experiments with $\tau \in \{0.1, 0.3, 0.5, 0.7\}$ and report the results in Table~\ref{tab:threshold}.
As shown, ThinkGR exhibits relatively stable performance across different threshold values, with the average recall ranging from 77.38\% to 78.44\%.
This robustness reduces the burden of extensive hyperparameter tuning in practice.
The optimal performance is achieved at $\tau = 0.5$, which we adopt in our main experiments.
A lower threshold (e.g., $\tau = 0.1$) classifies only completely failed retrievals as undesirable, providing insufficient negative supervision.
Conversely, a higher threshold (e.g., $\tau = 0.7$) may incorrectly label partially correct responses as undesirable, introducing noisy signals that hinder learning.
The moderate threshold $\tau = 0.5$ strikes a balance by ensuring clear differentiation between successful and failed retrievals, leading to a more robust and effective optimization signal.

\section{Index Storage Scalability Analysis}
\label{appendix:memory}

In addition to time efficiency, we also evaluate the index storage scalability of our method.
The FM-index used in ThinkGR is highly space-efficient.
As shown in Table~\ref{tab:memory}, our index is significantly smaller than GritHopper's dense retrieval index.
For instance, on the HotpotQA dataset with 5M documents, ThinkGR requires only 1.94 GB of index storage, which is 41$\times$ smaller than GritHopper's 80 GB.
This substantial reduction in index size makes ThinkGR highly feasible for large-scale retrieval scenarios involving hundreds of millions or even billions of documents.

\end{document}